\documentclass[12pt]{iopart}
\newcommand{\pT}{\mbox{$p_\perp$}}
\newcommand{\pTtrig}{\mbox{$\pT^{\mathrm{trig}}$}}
\newcommand{\pTassoc}{\mbox{$\pT^{\mathrm{assoc}}$}}
\newcommand{\ET}{\mbox{$E_\perp$}}
\newcommand{\zT}{\mbox{$z_\perp$}}

\newcommand{\eg}{\textit{e.g.}}
\newcommand{\vs}{\textit{vs.}}
\newcommand{\sqrts}{\mbox{$\sqrt{s}$}}

\newcommand{\upsi}{\mbox{$\Upsilon$}}
\newcommand{\gev}{\mbox{$\mathrm{GeV}$}}
\newcommand{\gevc}{\mbox{$\mathrm{GeV/c}$}}

\newcommand{\pp}{\mbox{$p+p$}}
\newcommand{\AuAu}{\mbox{$Au+Au$}}
\newcommand{\dAu}{\mbox{$d+Au$}}
\newcommand{\Npart}{\mbox{$N_{\mathrm{part}}$}}

\usepackage{iopams}
\usepackage{graphicx}
\usepackage{subfigure}
\usepackage{wrapfig}
\usepackage{cite}
\begin{document}

\title[STAR Highlights: High-$\pT$, EM \& Heavy Flavor Probes]{STAR Results on High Transverse Momentum, Heavy Flavor
and Electromagnetic Probes}

\author{M. Calder\'{o}n de la Barca S\'{a}nchez (for the STAR\footnote{Full collaboration list found in Appendix} Collaboration)}

\address{University of California, Davis\\
One Shields Ave\\
Davis, CA 95616} \ead{mcalderon@ucdavis.edu}
\begin{abstract}
We summarize here recent results from the STAR collaboration
focusing on processes involving large momentum transfers.
Measurements of angular correlations of di-hadrons are explored in
both the pseudorapidity ($\eta$) and azimuthal ($\phi$) projections.
In central \AuAu, an elongated structure is found in the $\eta$
projection which persists up to the highest measured \pT. After
quantifying the particle yield in this structure and subtracting it
from the near-side yield, we observe that the remainder exhibits a
behavior strikingly similar to that of the near-side yield in \dAu.
For heavy flavor production, using electron-hadron correlations in
\pp\ collisions, we obtain an estimate of the $b$-quark contribution
to the non-photonic electrons in the \pT\ region 3-6 \gevc, and find
it consistent with FONLL calculations. Together with the observed
suppression of non-photonic electrons in \AuAu, this strongly
suggests suppression of $b$-quark production in \AuAu\ collisions.
We discuss results on the mid-rapidity \upsi\ cross-section in \pp
collisions.  Finally, we present a proof-of-principle measurement of
photon-hadron ($\gamma-h$) correlations in \pp\ collisions, paving
the way for the tomographic study of the matter produced in central
\AuAu\ via $\gamma$-jet measurements.
\end{abstract}

\vspace{-1cm}
\vspace*{-0.5cm}
\section{Introduction}
Without question, some of the most exciting results that have been
observed in the study of relativistic heavy-ion collisions in the
RHIC era have been in the area of processes with large momentum
transfers.   All collaborations observed evidence of large
\emph{final state} suppression of the inclusive yield of charged
hadrons at high transverse momenta (\pT) in central \AuAu\
collisions
\cite{Back:2003ns,Adler:2003ii,Adams:2003im,Arsene:2003yk}, by as
much as a factor of 5.  The study of azimuthal correlations
demonstrated that the suppression mainly affects hadrons found in
the away-side of the leading particle in the event
\cite{Adams:2003im,Adler:2002tq}, and that this suppression was not
present in \dAu\ collisions. The consensus reached around these
observations was that RHIC produces extremely dense matter ($\sim
50$ times higher than ground-state nuclei), and the suppression
patterns were the result of induced gluon radiation as fast partons
traversed this dense medium
\cite{Arsene:2004fa,Back:2004je,Adams:2005dq,Adcox:2004mh}.

The focus for the last two years of RHIC research in this area has
been the quantitative determination of the properties of the
produced matter, \eg\ its density, temperature and viscosity.
Reliable estimations of these is the current challenge of the
heavy-ion program. To determine the density quantitatively, the
inclusive suppression was the first observable used to compare to
theory, but in essentially all theoretical descriptions the level of
suppression in the range $\lesssim 20 \gev$ shows a saturation with
increasing density, (see \eg \cite{Dainese:2004te,Eskola:2004cr}),
so further information is needed to constrain the models. Several
measurements can provide further clues. First, to complete a
systematic study of di-hadron correlations. Second, to use heavy
quarks as probes, given that they are expected to couple less
strongly to the medium. Finally, to measure photon-jet correlations
to vary the geometry bias compared to di-hadron correlations.
Together, the study of di-hadron correlations, heavy quarks and
$\gamma$-jet measurements should provide sufficient sensitivity to
quantitatively determine the density of the medium. We discuss now
the results relevant to these probes presented at this conference by
STAR. The summary of results for identified spectra at lower
momenta, strangeness, azimuthal anisotropy and three-particle
correlations is presented in Ref.~\cite{LJRuan:QM2006}.

\section{Di-hadron Correlations}

The interest in a systematic study of di-hadron correlations stems
from the need to better understand the interaction of fast partons
with the bulk of the produced medium. It is established that the
away-side particles are strongly modified by the dense matter,
leading to suppression of the yields \cite{Adler:2002tq}.  The
observation of broad away-side correlations at low
\pT\cite{Adams:2005ph} make clear that lowering the \pT\ is also
necessary to fully understand the system. Previous results on
di-hadron $\eta-\phi$ correlations used untriggered particles with
$\pT<2\ \gevc$\cite{Adams:2004pa}. At intermediate \pT, the
observation of dip-hump structures in the away-side correlation
functions \cite{Adler:2005ee,Ulery:QM2005} have generated much
interest and speculation as to the nature of the medium response to
a fast parton. At high \pT, clear di-jet signals are observed
\cite{Adams:2006yt} and the use of di-hadron fragmentation functions
has been proposed as an additional tool for quantitative comparison
to energy-loss phenomenology.  We now aim to map the angular
correlation panorama, extending the $\eta-\phi$ measurements using
triggered hadrons with $\pT>2\ \gevc$ \cite{JPutschke:QM2006},
forward-to- mid-rapidity $\phi$ correlations, \cite{LMolnar:QM2006},
and extending previous $\phi$ correlation and di-hadron
fragmentation measurements to study the \pTtrig\
systematics\cite{MJHorner:QM2006}.

We first summarize results from the $\Delta\phi-\Delta\eta$
correlation analysis at mid-rapidity ($|\eta|<1$). At low-\pT, there
are already quantitative measurements \cite{Adams:2004pa} of the
number and \pT\ correlation in $\eta-\phi$ which show a $\phi$
elongation of the 2-D number correlation in \pp\ collisions, and an
$\eta$ elongation (both in number and \pT\ correlations) in \AuAu\
for $\pT<2\ \gevc$. The results presented here extend the reach to
the region $2 < \pTassoc < 4\ \gev$ and $3< \pTtrig <12\ \gev$,
using the trigger-associated technique \cite{Adler:2002tq}.
\begin{figure}[htb]
  \begin{center}
\subfigure{\includegraphics[width=0.4\textwidth]{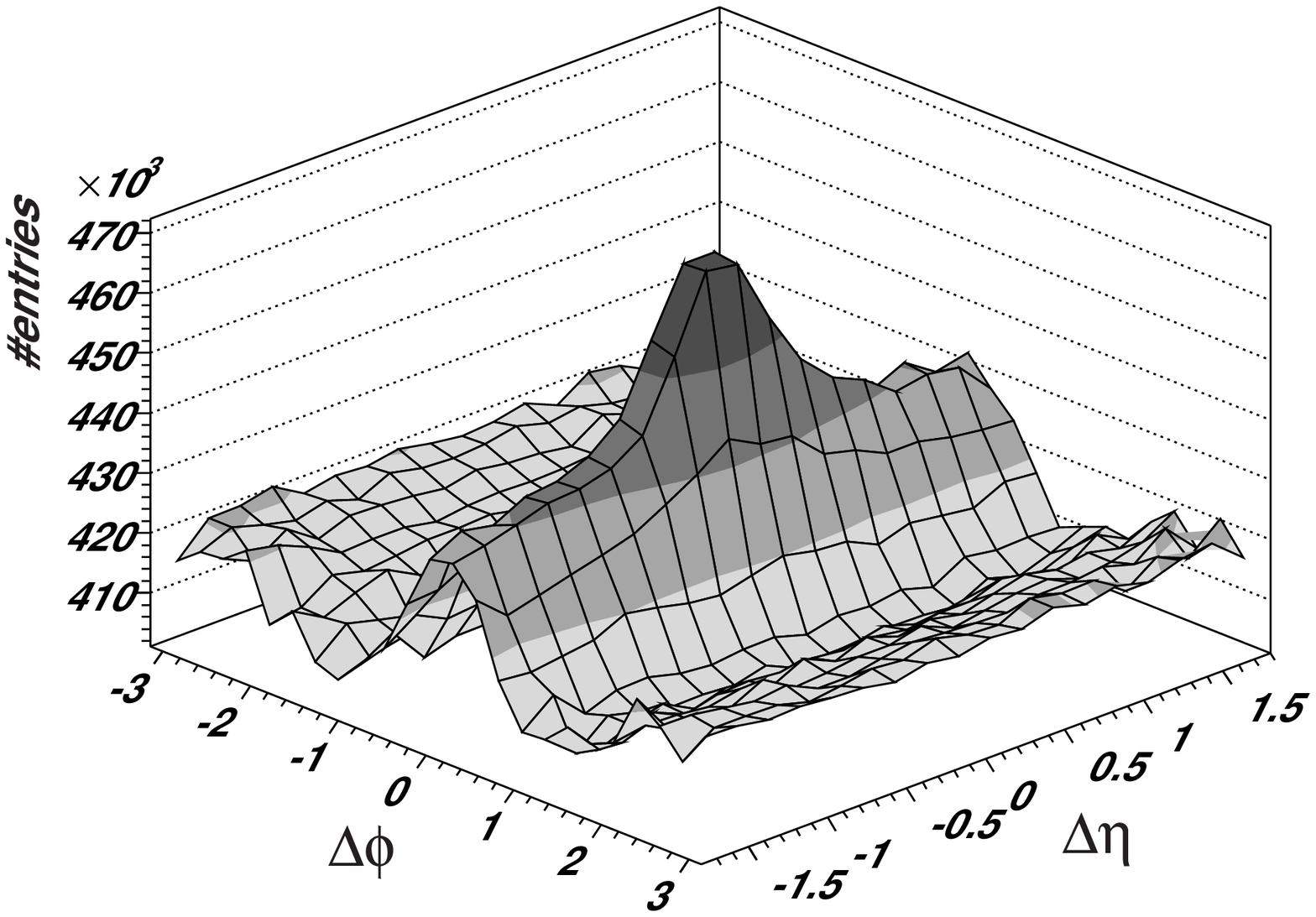}}
\subfigure{\includegraphics[width=0.4\textwidth]{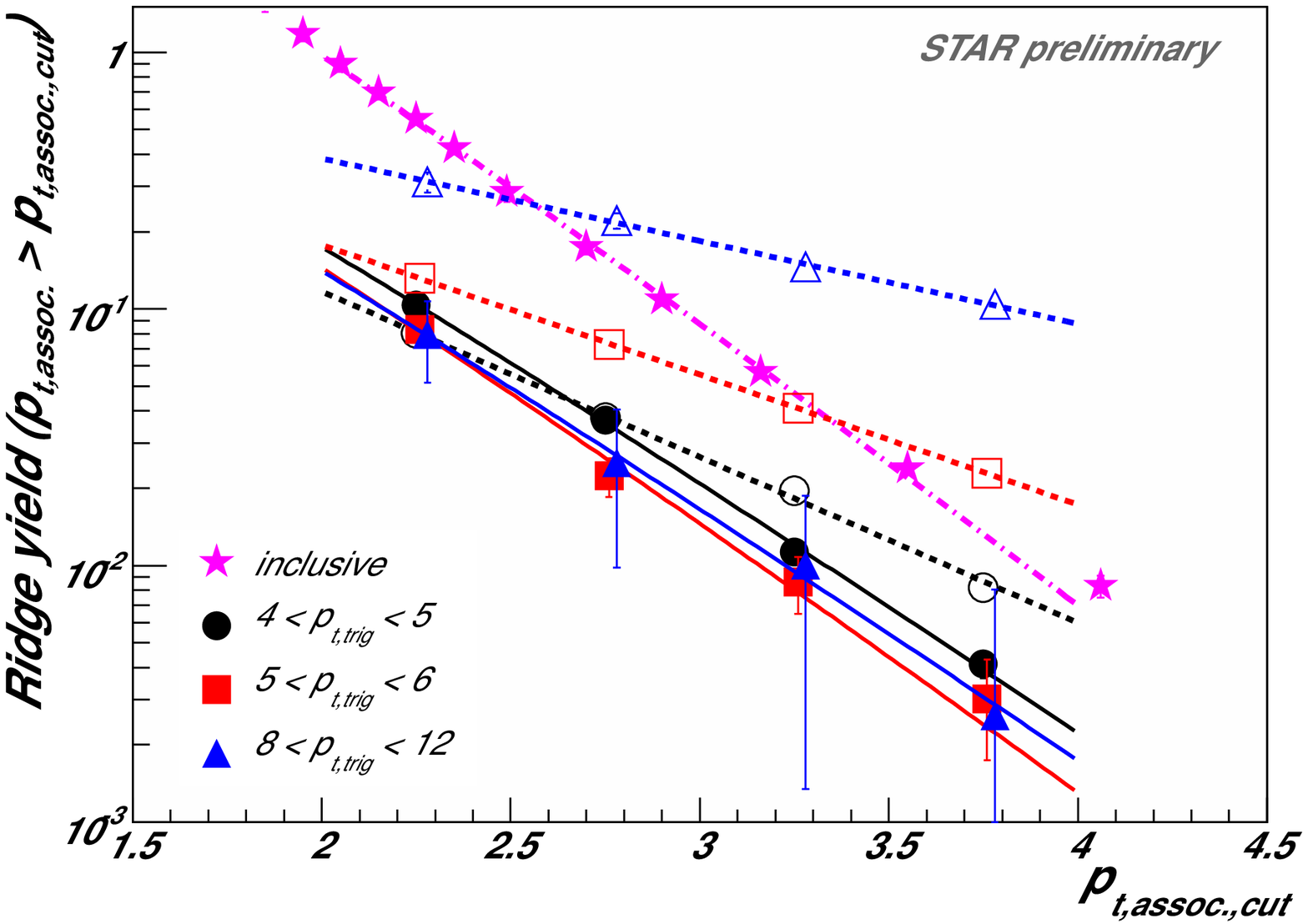}}
 \end{center}
   \caption[]{Left: $\Delta\eta-\Delta\phi$ correlation in \AuAu\ collisions,
   showing an elongation along the pseudorapidity-difference direction. We separate the correlation into two components which we label ``jet'' and ``ridge''
   (see text for details) and study the yields in each component. Right: Invariant yields \vs \pTassoc\ for the
   ``jet'' (dashed lines, open symbols) and ``ridge'' components (solid lines, filled symbols).
   The different symbols are for various choices of \pTtrig.}
  \label{fig:eta-phiCorrelations}
\end{figure}
Figure~\ref{fig:eta-phiCorrelations}(left), presents the
$\Delta\eta-\Delta\phi$ correlation, choosing tracks with $\pTassoc
>2$ \gev and $3<\pTtrig<4\ \gev$ in the 10\% most central \AuAu\
events.   We note the $v_2$ elliptic flow modulation in the
azimuth-difference axis, and for the region near $\Delta\phi=0$ an
$\eta$ correlation that is present throughout the $\eta$ acceptance
of the STAR TPC.  The correlation shows a maximum in the region near
$\Delta\eta=\Delta\phi=0$. To quantitatively study the yields as a
function of centrality and of the \pTassoc\ and \pTtrig\ kinematics,
we employed the following ansatz. We divide the correlation yield
into two components.  The first component is constructed to resemble
a jet-like correlation. We do this in three different ways. One is
to choose two disconnected regions in $\Delta\eta$, a region away
from the maximum ($0.7<|\Delta\eta|<1.4$) and a region near the
maximum ($|\Delta\eta|\leq 0.7$) (see
Ref.~\cite{JPutschke:QM2006,JBielcikova:QM2006} for more details and
examples). Subtracting the former from the latter, we obtain a
component that we label the ``jet''-like correlation
$\Delta\eta(J)$, borrowing from the picture of a QCD jet fragmenting
into hadrons. The yield remainder in the region $|\Delta\phi|<0.7$
we label the ``ridge''-like correlation, due to the obvious shape.
While this decomposition is instructive to produce quantitative
measurements of the yields \vs\ centrality and \vs\ the \pT\
kinematics, it should be noted that this does not imply that the
ridge-like correlation is unrelated to jet-fragmentation phenomena.

Once we use the decomposition into the ``jet'' ($J$) and the
``ridge'' ($R$), we can study the yields.  We find the total yield
($J+R$) increases linearly as a function of \Npart, but after
removing the ridge yield $R$, the remainder $J$ yield is independent
of centrality (see \cite{JPutschke:QM2006}).  Focusing on central
collisions, Fig.\ref{fig:eta-phiCorrelations}(right) shows the yield
in varying bins of \pTassoc\ for windows of \pTtrig\ for the $J$
component (open symbols) and the $R$ component (filled symbols).  We
fit an exponential to each resulting \pTassoc\ spectrum; dashed
lines for the $J$ spectra and solid lines for the $R$ spectra. For
reference, the inclusive spectrum is shown (stars). We observe that
the $R$ spectra have similar slopes to the inclusive spectra, and
the overall shape and yields are independent of $\pTtrig$ within our
uncertainties. In contrast, the $J$ spectra are all harder than the
inclusive, and a clear correlation is seen between the slope and
yield of the $J$ spectrum and the choice of \pTtrig.
\begin{wrapfigure}[20]{r}[0.01\textwidth]{0pt}
\includegraphics[width=0.7\textwidth]{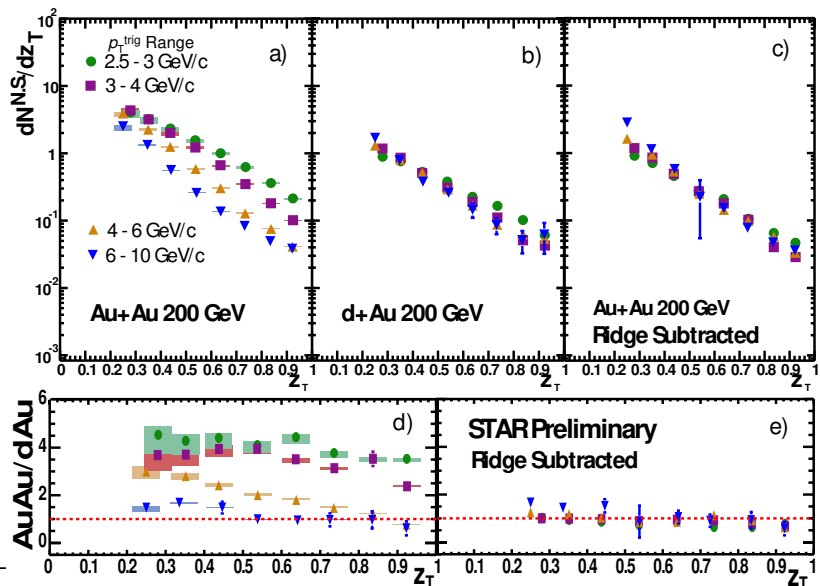}
   \caption[]{\AuAu\ (a) and \dAu\ (b) \zT\ distributions.  The ratio of \AuAu\ to \dAu\ is shown in (d).
   \AuAu\ after subtraction of the ``ridge''
   yield is shown in (c), and its ratio to \dAu\ is shown in (e).}
   \label{fig:zTDistributions}
\end{wrapfigure}
Studies on $\Delta\phi$ correlations can now benefit from the
knowledge of this elongated $\Delta\eta$ structure in the near-side.
Figure \ref{fig:zTDistributions} shows the result of applying the
$J+R$ ansatz to the near-side di-hadron fragmentation functions \vs\
$\zT\equiv\pTassoc/\pTtrig$.

Figure~\ref{fig:zTDistributions}a) shows the \AuAu\ \pT-triggered
di-hadron fragmentation functions \vs\ \zT\ for various windows in
\pTtrig. The top-middle panel shows the same study for \dAu, where
the \zT\ dependence of the \dAu\ data has a similar shape for all
choices of \pTtrig.  Panel d) shows the ratio of the \zT\ spectra in
\AuAu\ to that in \dAu.  All \zT\ spectra have similar slopes in
\AuAu\ compared to \dAu, but the yields are larger in \AuAu\ for
lower \pTtrig.  Removing the $R$ component, as shown in
Fig.~\ref{fig:zTDistributions}c), we find the remaining particles
behave just like \dAu: Fig.~\ref{fig:zTDistributions}e) shows the
ratio of the $R$ removed \zT\ spectra in \AuAu\ to the \dAu\
spectrum, seen to be close to unity within our uncertainties for all
\pTtrig\ windows.  This is suggestive that the $J$ component might
be indicative of vacuum fragmentation after energy loss. Together
with the quantitative study of the low-\pT\ number and \pT\
correlations, these measurements should provide a reference against
which models attempting to constrain the density of the medium can
be compared.

\section{Heavy Flavor}

\subsection{e-h Correlations}
The trigger-associated correlation technique has been extended to
use electron triggers. Experimentally, measurements of the
non-photonic electron spectra are used as a proxy for heavy quarks.
These show a rather large amount of suppression\cite{Abelev:2006db}
in \AuAu, much larger than initial calculations of heavy-quark
energy loss predicted\cite{Dokshitzer:2001zm}. The mechanism of
energy loss for heavy quarks was reassessed, with the conclusion
that collisional energy loss is important\cite{Djordjevic:2006tw}.
\begin{wrapfigure}[16]{r}[0.01\textwidth]{0pt}
\vspace*{-.5cm}
  \includegraphics[width=0.5\textwidth]{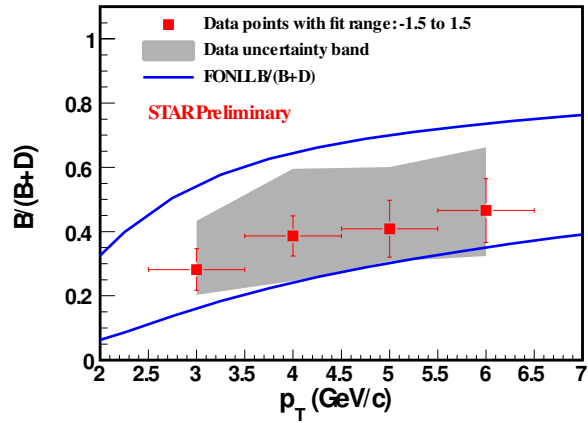}
   \caption{B-meson contribution to the non-photonic electrons extracted from
   a fit to $e-h$ azimuthal correlations in $\pp$ collisions.}
  \label{fig:bContribution}
\end{wrapfigure}
Nevertheless, the calculations were still not able to reproduce the
level of suppression measured for the non-photonic electrons.  Since
the measurement is sensitive to the sum of the $c$ and $b$ quark
contributions, it was also noted that if the $b$-quark contribution
was modified (and in the extreme case, \emph{ignored}) the
calculations were consistent with the measurement.  Therefore, an
urgent need arose for disentangling the relative contributions of
$c$ and $b$ quarks to resolve the ambiguity.  We present here the
first results from a method to extract the $b$ contribution to the
non-photonic electrons via the analysis of $e-h$ correlations in
\pp\ collisions.  The method exploits the correlation between the
electron and hadron that originate from the semi-leptonic decay of a
heavy-quark. The larger mass of the $B$ meson results in a larger
energy given to the decay products, which results in a broader
near-side $e-h$ correlation compared to that resulting from the
semi-leptonic decay of a $D$ meson.  We have simulated the resulting
$e-h$ correlation shapes using PYTHIA, which introduces a
model-dependence into the method. The model-dependence is small
however: varying the underlying fragmentation functions we find no
significant modification of the $B$ and $D$ correlation shape. In
other words, the correlation shapes are to first order sensitive
only to the kinematics of the decay given the meson masses (a much
tighter constraint). The details of the analysis (background
subtraction, fitting procedure and $B/(B+D)$ ratio extraction) are
in Ref.~\cite{XYLin:QM2006}. Figure~\ref{fig:bContribution} shows
the resulting $B/(B+D)$ ratio from this analysis (filled squares).
The error bars are statistical, and the shaded band is the
systematic uncertainty (mainly from photonic-background
reconstruction efficiency).  The two lines in the figure are from
FONLL calculations, showing the bounds of the theoretical
uncertainties (mass and renormalization scale).  The data from this
measurement are consistent with the FONLL calculations.  This has
important implications for the \AuAu\ data. Since model calculations
attempting to describe the non-photonic electron suppression in
\AuAu\ were not able to reproduce the large amount of suppression, a
better agreement could be arrived at assuming the $B$ contribution
only starts to appear at $\sim 10\ \gevc$.  The data in
Fig.~\ref{fig:bContribution} disallows such solution, and one is
left with the conclusion that to explain the data it is necessary to
suppress the $B$ mesons in central \AuAu\ at RHIC.  A possible
mechanism for $B$ suppression discussed in Ref.~\cite{Adil:2006ra}
invokes collisional dissociation of the $B$ mesons.

\subsection{\upsi Production}\label{sec:upsilon}
\begin{figure}[htb]
  \begin{center}
\subfigure{\includegraphics[width=0.5\textwidth]{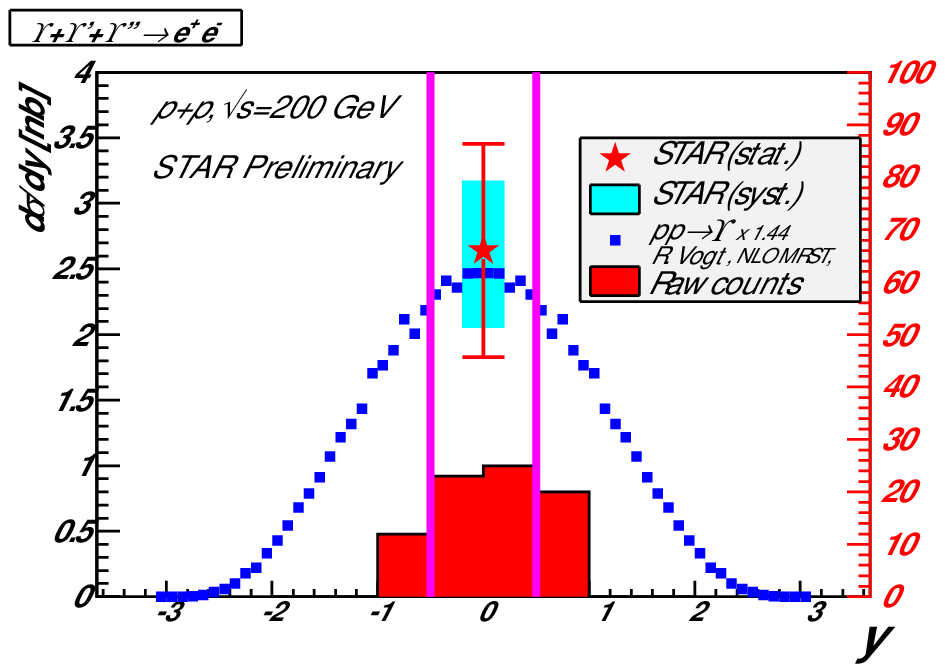}}
\subfigure{\includegraphics[width=0.3\textwidth]{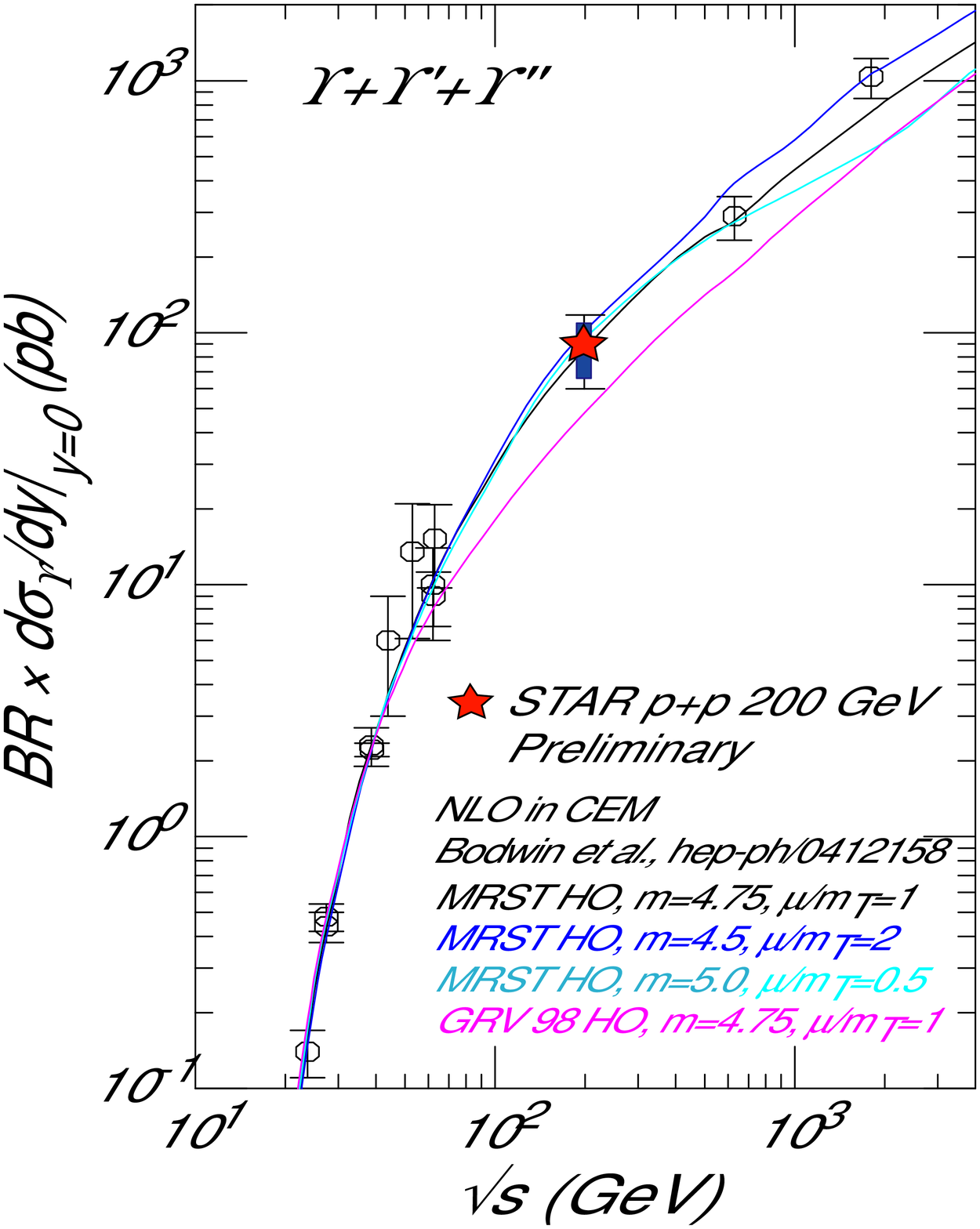}}
 \end{center}
   \caption[]{Left: $\upsi$ cross section ((1s+2s+3s), filled star) measured in $|y|<0.5$ (see left abscissa for
   scale) compared to NLO calculations.
   The raw yield \vs\ $y$ is shown as the filled histogram (see right abscissa for scale).  Right:
Comparison of the
  measured cross section to world data and NLO calculations. See text for details.}
  \label{fig:upsilon}
\end{figure}
During 2006, several analyses benefited from the completion of the
STAR EMC ($|\eta|<1$ and $0<\phi<2\pi$). One is a measurement of the
\upsi\ production at mid-rapidity. The larger coverage of the
calorimeter improves the \upsi\ acceptance in STAR by a factor of 4
over the coverage during the 2004 run. During the 2006 \pp\ run, we
sampled an integrated luminosity of $\mathcal{L}=9\
\mathrm{pb}^{-1}$.  Two main trigger settings were used during the
running period; we analyzed and obtained full corrections for one of
these. Figure~\ref{fig:upsilon}(left) shows the measurement of the
\upsi(1s+2s+3s) cross section $d\sigma/dy|_{y=0}$. The filled
histogram shows the background-subtracted counts in the \upsi\ mass
region to illustrate the acceptance. The left abscissa shows the
$d\sigma/dy$ scale and the right abscissa shows the scale for the
raw-count histogram. The error bars in the cross section measurement
are the statistical error and the shaded rectangle is the systematic
uncertainty. The low cross section of the \upsi\ at RHIC makes this
a luminosity limited measurement.  We compare the datum point to a
NLO calculation of the $\upsi$(1s) cross section scaled by 1.44
(based on CDF \cite{Abe:1995an}) to account for the excited states.
Figure~\ref{fig:upsilon}(right) shows the the world data on \upsi\
production as a function of $\sqrts$, compared to NLO calculations.
The NLO calculation is consistent with our measurement within our
uncertainties.  Future measurements of \upsi\ production in \AuAu\
will yield additional clues as to the temperature of the produced
matter.

\section{$\gamma-h$ Correlations}\label{sec:gamma-h}
\begin{figure}[htb]
\subfigure{\includegraphics[width=0.55\textwidth]{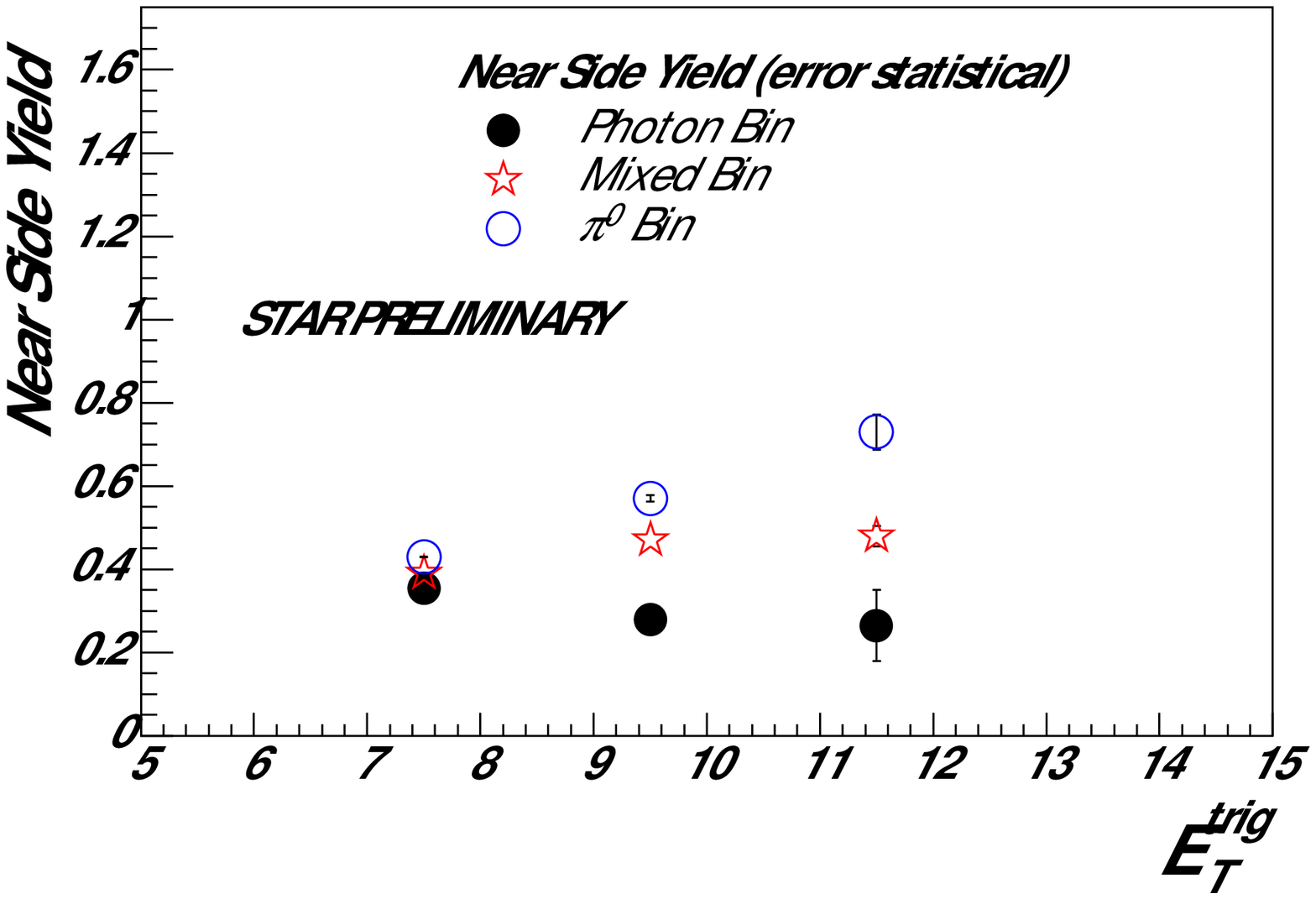}}
\subfigure{\includegraphics[width=0.4\textwidth]{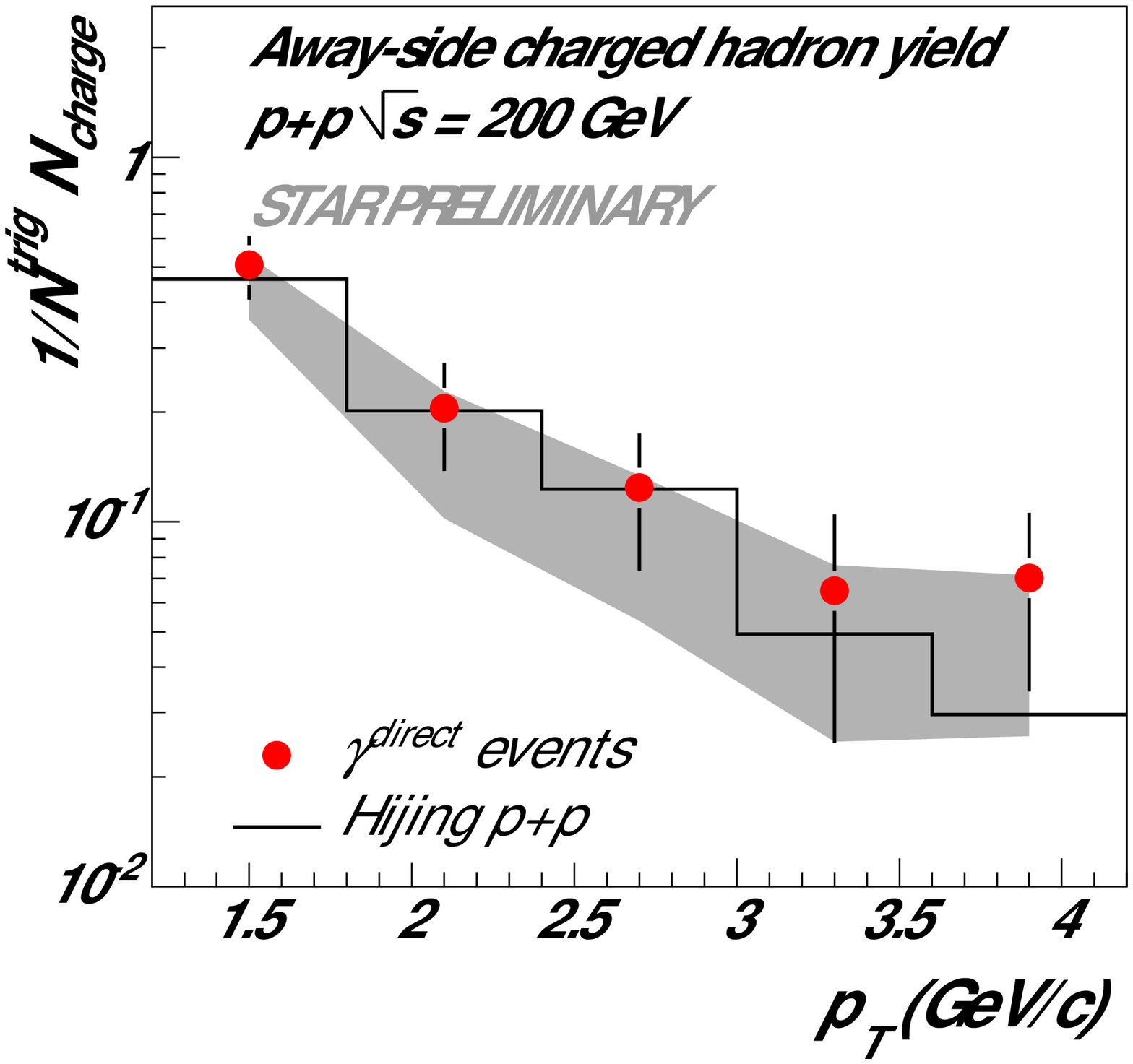}}
  \caption[]{Left: Integrated yield of the near-side $\Delta\phi$
  correlation in $\pp$ collisions
  for the $\pi^0$-enriched (open circles), mixed (open stars), and $\gamma$-enriched
  (filled circles) samples \vs\ $\ET$.
  Right: For the $\gamma$-enriched sample,
  the measured away-side yield of charged hadrons (circles) is compared to HIJING
  simulations selecting events with direct $\gamma$ production.}
  \label{fig:gamma-hVsEt}
\end{figure}
Finally, we present results from the analysis of photon-hadron
($\gamma-h$) correlations in \pp\ collisions.  This analysis has
also benefited from the increased coverage of the calorimeter.
Compared to two-particle correlations using di-hadrons, $\gamma-h$
measurements in \AuAu\ hold the promise to essentially remove the
surface bias that complicates the interpretation and model
comparison of the correlations. Since the photon will not suffer the
effects of energy loss it can be used as a reference for the
momentum transfer of the partonic hard process.

In the analysis of $\gamma-h$ correlations, the dominant source of
background to the measurement of direct photons is $\pi^0$
contamination.  The EMC shower maximum detector (SMD) is employed as
a discriminator between direct $\gamma$'s and $\pi^0$'s (for
details, see Ref.~\cite{SChattopadhyay:QM2006}). The SMD showers can
be used to select showers with $\pi^0$-like shape, $\gamma$-like
shape, or somewhere in between. The tell-tale signature of a direct
$\gamma$ is that it should have no associated hadrons near it in
$\eta-\phi$.  Figure~\ref{fig:gamma-hVsEt}(left) shows the near-side
yields of the $\pi^0$-enriched (open circles), mixed (open stars),
and $\gamma$-enriched (filled circles) samples \vs\ $\ET$. The
$\gamma$-enriched data show a decrease of the yield at high \ET,
consistent with the expected behavior for $\gamma$-jet events.  For
this sample, the away-side yield is compared to $\gamma$-jet
simulations using HIJING in Fig.~\ref{fig:gamma-hVsEt}(right).  The
observed \pT\ spectrum of the away side yields is consistent with
the calculations within our experimental uncertainty, an encouraging
development for future tomographic explorations of the medium using
high-\ET\ photons.

\section{Conclusions}
We have measured di-hadron correlations in a large fraction of the
available kinematics at RHIC energies.  We presented in this
conference results on $\Delta\eta-\Delta\phi$ correlations at
intermediate- to high-\pT.  Using a simple ansatz to isolate the
yield in the jet-like near-side region and subtracting it from the
total to quantify the yield in the ``ridge'' region, we observe that
the near-side jet-like peak shows compatibility with vacuum
fragmentation behavior.  Our hope is that with the quantitative
measurements of the correlations from low, intermediate and
high-\pT\ will be less ``fragile'' than single inclusive spectra,
and this robustness can provide tighter constraints on the
estimations of the density. The analysis of $e-h$ correlations has
yielded the first estimate of the $B$ contribution to the
non-photonic electrons in \pp; we find it consistent with FONLL
calculations. Together with the measured suppression of the
non-photonic electron inclusive spectrum, the corollary of this
observation is that there must be $B$ suppression at RHIC in \AuAu.
We presented first \pp\ results from two luminosity-hungry
measurements: the \upsi\ cross section and the $\gamma-h$
correlation. A larger integrated luminosity in future runs and RHIC
II should allow a measurement of the \upsi\ (2s)/(1s) and (3s)/(1s)
ratios: perhaps one of the most direct ways to connect experimental
data to a temperature calculated from lattice QCD.  The $\gamma-h$
measurement allows experimentalists to vary the surface bias that
exists in measurements relying on particles which suffer energy
loss, so it can be one of the ultimate probes of the medium density.
Having performed the proof-of-principle measurements, we now eagerly
await a chance to take the data to complete these programs.


\section{References}
\begin{thebibliography}{9}
\bibitem{Back:2003ns}
  B.~B.~Back {\it et al.}  [PHOBOS Collaboration],
  Phys.\ Rev.\ Lett.\  {\bf 91}, 072302 (2003)

\bibitem{Adler:2003ii}
  S.~S.~Adler {\it et al.}  [PHENIX Collaboration],
  Phys.\ Rev.\ Lett.\  {\bf 91}, 072303 (2003)

\bibitem{Adams:2003im}
  J.~Adams {\it et al.}  [STAR Collaboration],
  Phys.\ Rev.\ Lett.\  {\bf 91}, 072304 (2003)

\bibitem{Arsene:2003yk}
  I.~Arsene {\it et al.}  [BRAHMS Collaboration],
  Phys.\ Rev.\ Lett.\  {\bf 91}, 072305 (2003)

\bibitem{Adler:2002tq}
  C.~Adler {\it et al.}  [STAR Collaboration],
  Phys.\ Rev.\ Lett.\  {\bf 90}, 082302 (2003)

\bibitem{Arsene:2004fa}
  I.~Arsene {\it et al.}  [BRAHMS Collaboration],
  Nucl.\ Phys.\ A {\bf 757}, 1 (2005)

\bibitem{Back:2004je}
  B.~B.~Back {\it et al.},
  Nucl.\ Phys.\ A {\bf 757}, 28 (2005)

\bibitem{Adams:2005dq}
  J.~Adams {\it et al.}  [STAR Collaboration],
  Nucl.\ Phys.\ A {\bf 757}, 102 (2005)

\bibitem{Adcox:2004mh}
  K.~Adcox {\it et al.}  [PHENIX Collaboration],
  Nucl.\ Phys.\ A {\bf 757}, 184 (2005)

\bibitem{Dainese:2004te}
  A.~Dainese, C.~Loizides and G.~Paic,
  Eur.\ Phys.\ J.\ C {\bf 38}, 461 (2005)

\bibitem{Eskola:2004cr}
  K.~J.~Eskola, H.~Honkanen, C.~A.~Salgado and U.~A.~Wiedemann,
  Nucl.\ Phys.\ A {\bf 747}, 511 (2005)

\bibitem{LJRuan:QM2006}
 L.~J.~Ruan, {\it et al.}  [STAR Collaboration],
 these proceedings.

\bibitem{Adams:2005ph}
  J.~Adams {\it et al.}  [STAR Collaboration],
  Phys.\ Rev.\ Lett.\  {\bf 95}, 152301 (2005)

\bibitem{Adams:2006yt}
  J.~Adams {\it et al.}  [STAR Collaboration],
  Phys.\ Rev.\ Lett.\  {\bf 97}, 162301 (2006)

\bibitem{Adler:2005ee}
  S.~S.~Adler {\it et al.}  [PHENIX Collaboration],
  Phys.\ Rev.\ Lett.\  {\bf 97}, 052301 (2006)

\bibitem{Ulery:QM2005}
  J.~G.~Ulery {\it et al.}  [STAR Collaboration],
  Nucl.\ Phys.\ A {\bf 774}, 581 (2006)

\bibitem{JPutschke:QM2006}
  J.~Putschke {\it et al.} [STAR Collaboration],
  these proceedings.

\bibitem{LMolnar:QM2006}
  L.~Molnar {\it et al.} [STAR Collaboration],
  these proceedings.

\bibitem{MJHorner:QM2006}
  M.~J.~Horner {\it et al.} [STAR Collaboration],
  these proceedings.

\bibitem{Adams:2004pa}
  J.~Adams {\it et al.}  [STAR Collaboration],
  Phys.\ Rev.\ C {\bf 73}, 064907 (2006)

\bibitem{JBielcikova:QM2006}
  J.~Bielcikova {\it et al.} [STAR Collaboration],
  these proceedings.

\bibitem{Abelev:2006db}
  B.~I.~Abelev {\it et al.}  [STAR Collaboration],
  arXiv:nucl-ex/0607012.

\bibitem{Dokshitzer:2001zm}
  Y.~L.~Dokshitzer and D.~E.~Kharzeev,
  Phys.\ Lett.\ B {\bf 519}, 199 (2001)
  [arXiv:hep-ph/0106202].

\bibitem{Djordjevic:2006tw}
  M.~Djordjevic,
  Phys.\ Rev.\ C {\bf 74}, 064907 (2006)
  [arXiv:nucl-th/0603066].

\bibitem{XYLin:QM2006}
  X.~Y.~Lin {\it et al.} [STAR Collaboration],
  these proceedings.

\bibitem{Adil:2006ra}
  A.~Adil and I.~Vitev,
  arXiv:hep-ph/0611109.

\bibitem{Abe:1995an}
  F.~Abe {\it et al.}  [CDF Collaboration],
  Phys.\ Rev.\ Lett.\  {\bf 75}, 4358 (1995).
\bibitem{SChattopadhyay:QM2006}
  S.~Chattopadhyay {\it et al.} [STAR Collaboration],
  these proceedings.
  \end{thebibliography}
\end{document}